\documentclass[conference]{IEEEtran}

\usepackage[ruled]{algorithm2e}

\SetAlFnt{\small}
\SetAlCapFnt{\small}
\SetAlCapNameFnt{\small}
\SetAlCapHSkip{0pt}
\IncMargin{-\parindent}

\usepackage{tikz}
\usepackage{pgf}
\usepackage{amssymb}
\usetikzlibrary{arrows,positioning,calc,fit,backgrounds,shapes,automata,,decorations,}
\usepackage{pgfplots,pgfplotstable} 
\usepgfplotslibrary{units}
\usetikzlibrary{external} 
\usetikzlibrary{spy,shapes.misc,shapes.geometric}
\usepgfplotslibrary{groupplots,external}

\usepackage{enumitem}

\usepackage{array}
\usepackage{multirow}
\usepackage{listings}
\usepackage{xspace}

\usepackage{mathrsfs}
\usepackage{filecontents}

\usepackage{amsmath}
\usepackage{amsfonts}
\usepackage{amssymb}
\usepackage{calligra}
\usepackage{calrsfs}

\usepackage[mathscr]{euscript}

\newtheorem{lemma}{Lemma}

\usepackage{acronym}
\newcommand{\TsToCam}{\Delta_{s_1 \rightarrow a_1}}
\newcommand{\conSym}{\mathcal{C}}
\newcommand{\obsSym}{\mathcal{O}}

\newcommand{\realN}{ \mathbb{R}\xspace} 
\newcommand{\noAssumptions}{ \top\xspace } 
\newcommand{\within}{\xspace\textbf{within}\xspace} 
\newcommand{\every}{\xspace\textbf{every}\xspace} 

\newcommand{\ignore}[1]{{}}
\newcommand{\sid}[1]{{#1}}

\newcolumntype{L}[1]{>{\raggedright\let\newline\\\arraybackslash\hspace{0pt}}m{#1}}
\newcolumntype{C}[1]{>{\centering\let\newline\\\arraybackslash\hspace{0pt}}m{#1}}
\newcolumntype{R}[1]{>{\raggedleft\let\newline\\\arraybackslash\hspace{0pt}}m{#1}}

\definecolor{graphFirst}{RGB}{2,136,209} 
\definecolor{graphSecond}{RGB}{211,47,47} 
\definecolor{graphThird}{RGB}{245,124,0} 
\definecolor{graphFourth}{RGB}{56,142,60} 
\definecolor{graphFifth}{RGB}{81,45,168} 
\definecolor{graphSixth}{RGB}{69,90,100} 
\definecolor{graphSeventh}{RGB}{251,192,45} 

\definecolor{backgroundFirst}{RGB}{129,212,250} 
\definecolor{backgroundSecond}{RGB}{239,154,154} 
\definecolor{backgroundThird}{RGB}{255,204,128} 
\definecolor{backgroundFourth}{RGB}{165,214,167} 
\definecolor{backgroundFifth}{RGB}{179,157,219} 
\definecolor{backgroundSixth}{RGB}{176,190,197} 
\definecolor{backgroundSeventh}{RGB}{255,245,157} 

\usepackage{xargs}   
\usepackage[colorinlistoftodos,prependcaption,textsize=tiny]{todonotes}
\newcommandx{\unsure}[2][1=]{\todo[linecolor=red,backgroundcolor=red!25,bordercolor=red,#1]{#2}}
\newcommandx{\change}[2][1=]{\todo[linecolor=blue,backgroundcolor=blue!25,bordercolor=blue,#1]{#2}}
\newcommandx{\info}[2][1=]{\todo[linecolor=OliveGreen,backgroundcolor=OliveGreen!25,bordercolor=OliveGreen,#1]{#2}}
\newcommandx{\improvement}[2][1=]{\todo[linecolor=Plum,backgroundcolor=Plum!25,bordercolor=Plum,#1]{#2}}
\newcommandx{\thiswillnotshow}[2][1=]{\todo[disable,#1]{#2}}

\IEEEoverridecommandlockouts

\begin{document}
	

	\acrodef{ALM}{Adaptive Logic Module}
\acrodef{APD}{Action Potential Duration}
\acrodef{BCL}{Base Cycle Length}
\acrodef{CPS}{Cyber-Physical System}
\acrodef{DI}{Diastolic Interval}
\acrodef{DSP}{Digital Signal Processor}
\acrodef{DTTS}{Discrete Time Transition System}
\acrodef{DHA}{Deterministic Hybrid Automata}
\acrodef{EA}{Evolutionary Algorithm}
\acrodef{ECG}{Electrocardiogram}
\acrodef{EGM}{Electrogram}
\acrodef{FPGA}{Field Programmable Gate Arrays}
\acrodef{HA}{Hybrid Automata}
\acrodef{HIOA}{Hybrid Input Output Automata}
\acrodef{ILP}{Integer Linear Programming}
\acrodef{MCU}{Microcontroller Unit}
\acrodef{ODE}{Ordinary Differential Equation}
\acrodef{PoC}{Plant-on-a-Chip}
\acrodef{QoS}{Quality of Service}
\acrodef{SHIOA}{Synchronous Hybrid Input Output Automata}
\acrodef{SWA}{Synchronous Witness Automata}
\acrodef{TA}{Timed Automata}
\acrodef{WHA}{Well-formed Hybrid Automata}
\acrodef{WCET}{Worst-Case Execution Time}
\acrodef{FSM}{Finite State Machine}

\acrodef{ST}{Stimulated}
\acrodef{UP}{Upstroke}
\acrodef{ERP}{Effective Refractory Period}
\acrodef{RRP}{Relative Refractory Period}
\acrodef{RP}{Resting Period}
\acrodef{AP}{Action Potential}

\acrodef{SA}{Sinoatrial}
\acrodef{AV}{Atrioventicular}
\acrodef{RVA}{Right Ventricular Apex}

\acrodef{QSS}{Quantized State System}

	
	
	\title{\huge Contract-based Methodology for Developing \\ Resilient Cyber-Infrastructure in the Industry 4.0 Era\vspace{-0.4cm}}
	
	\author{
	
	\IEEEauthorblockN{Sidharta Andalam, Daniel  Jun Xian Ng, Arvind Easwaran}
	\IEEEauthorblockA{Nanyang Technological University (NTU), 
		Singapore\\
		Email: arvinde@ntu.edu.sg}
	\and
	\IEEEauthorblockN{Karthikeyan Thangamariappan}
	\IEEEauthorblockA{Delta Electronics Inc.,
		Singapore\\
		Email: karthikeyan.t@deltaww.com}

\thanks{
	This work was conducted within the Delta-NTU Corporate Lab for Cyber-Physical Systems with funding support from Delta Electronics Inc. and the National Research Foundation (NRF) Singapore under the Corp Lab@University Scheme.
} 

	}
	
	\maketitle
		
		\IEEEpubid{
	} 
		
	\begin{abstract}



As the  industrial cyber-infrastructure become increasingly important to realise the objectives of Industry~4.0, the consequence of disruption 
 due to internal or external faults
  become increasingly severe. Thus there is a need for a resilient infrastructure.
 In this paper,
we propose a contract-based methodology
where components across layers of the cyber-infrastructure are associated with \emph{contracts} and a light-weight resilience manager.  This allows the system to detect faults (contract violation monitored using observers) and react (change contracts dynamically) effectively. 

\end{abstract}



	\maketitle

	\section{Introduction}
	\label{sec:introduction}


Industry 4.0 has the potential to radically improve the 
 productivity of manufacturing systems. 
The next generation of smart factories 
 can perform more efficiently, collectively and resiliently~\cite{LEE201518, Vyatkin2017, Sztipanovits2012}.
A resilient system has the ability to maintain and improve
services even when challenged by failures and evolutionary changes. It indicates a system to be more flexible, more dynamic and less prescriptive than a traditional fault tolerant system~\cite{Strigini2012}.


Resiliency can be obtained using software or hardware based fault tolerance mechanisms.
The latter solution demands physical redundancy such as replication
of computational and communicational resources.
For software based solutions, we find techniques such as 
replication of programs, checkpoints that act as restoration points and 
monitoring services that rely  on timestamps~\cite{Eisele2017}  and
fault trees, to detect faults.
More recently, we see software adaptation techniques 
that switch between off-line generated code based on ontology~\cite{OliverH2017} or application graphs~\cite{Valls2013}, or 
dynamically create/modify code at runtime~\cite{Vyatkin2017,Lepuschitz2011}.

Our cyber-infrastructure architecture follows the classical layered middle-ware with three different layers~\cite{Sztipanovits2012}, see Fig.~\ref{fig:CILayers}.
The physical layer comprises physical components such as sensors, actuators, controllers and communication hardware. 
The platform layer embodies computational and communicational platforms such as operating systems and network managers.  
Finally, the application layer accommodates the software components which describe the behaviour of an application. 

In~\cite{Eisele2017} and \cite{Denker2012}, authors discuss an outline for a holistic approach for developing a resilient cyber-infrastructure that manages applications, devices, resources, deployment, security and time etc. 
However,~\cite{Denker2012} does not present any architecture details. In contrast, RIAPS~\cite{Eisele2017}
 describes an architecture for developing a distributed resilient CPS. As an example,
a resilient discovery service (DS)  was explored; a failure of a publisher/subscriber pair is detected using heartbeat signals and timestamps. 
When a failure occurs, DS de-registers the pair from the
list of registered services. A publisher/subscriber needs to re-register once they become active.
The process of de-registering and re-registering is very time consuming and it also needs to be
communicated with neighbouring nodes. In this scenario, the cause of the failure (e.g. intermittent fault in the physical layer) and the expected recovery time (e.g. 2 seconds or 2 hours) were not
available to the DS manager residing in the platform layer.  If the recovery time
information was available to the DS manager, it can choose not to de-register a publisher/subscriber
and avoid the unnecessary time consuming registration process across all neighbouring nodes. Thus, there is a need for communication across layers to improve resiliency.



\subsection{Problem statement} Existing resiliency techniques are inefficient due to two key limitations: {\color{black} (1)  They do not consider cross-layer interactions of a cyber infrastructure~\cite{Vyatkin2017,OliverH2017,Eisele2017,Valls2013,Lepuschitz2011}. (2) The techniques depend on a centralised decision-making component which collects information from other  distributed components to detect faults~\cite{OliverH2017,Valls2013}, except some  which are tightly coupled  to an industry standard~\cite{Vyatkin2017,Lepuschitz2011}.
 Both limitations (\emph{lack of cross-layer communication} and \emph{centralised decision making}) are not acceptable for Industry~4.0 where the cyber-infrastructure  is heavily distributed and  encourages attributes such as \emph{self-awareness} where even the edge nodes of the network are expected to make decisions~\cite{LEE201518, Vyatkin2017}.
 This is unlike most existing pyramid-like infrastructure with predefined  hierarchy of connections from sensors/actuators at the lowest levels and decision making software at the highest levels.
Hence, there is a need for a new resilient methodology  that allows for a more fine-grain distributed resilience management with cross-layer interactions and is aligned with the attributes of Industry~4.0. 


\subsection{Proposed solution}

\begin{figure}[tbhp]
	\centering
	\includegraphics[width=0.95\columnwidth]{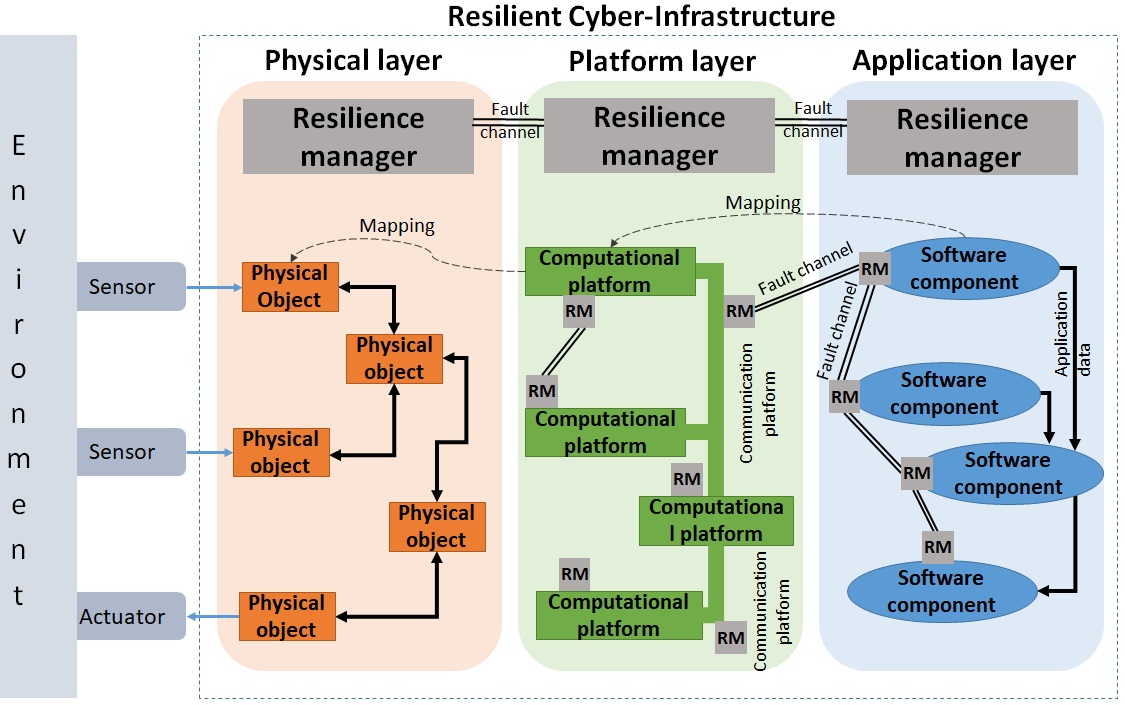}
	\caption{\label{fig:CILayers} Proposed contract-based methodology for resiliency. Its shows a distributed resiliency management (RM) at component-level  and across layers. 
		\vspace{-0.5cm}}
\end{figure}

In this paper, we propose contract-based methodology
where components across layers of the cyber-infrastructure are associated with \emph{contracts} and a light-weight resilience manager, see Fig.~\ref{fig:CILayers}.  This allows the system to detect faults (contract violation detected using Timed or Hybrid Automata) and react (change contracts dynamically) effectively.
\sid{When a component-level resilience manager is unable  to provide any feasible solution to a fault, a layer-level resilience manager is informed. It has global-level understanding of the system and may find a feasible solution.  }
We now discuss how the prosed methodology aligns with the attributes of Industry 4.0. 


\textbf{Self-reconfigurability}: This refers to the ability for the individual node (or a component) to detect disturbances and apply corrective and pre-emptive measures. This attribute aligns with the core contribution of the proposed methodology. We use contracts to clearly define the functionality of a component. Any disturbances that violate the contracts are monitored using observers that run concurrently. As a recovery strategy
a component switches between contracts  (changing it's behaviour) to contain faults, a step towards ensuring zero downtime production.  

\textbf{Self-optimisation}:  The availability of the resources in a cyber-infrastructure changes dynamically, {\color{black} detected using a discovery service~\cite{Eisele2017}.} The nodes are expected to self-optimise to improve performance.  In order to maximise the
performance, component manager  tries to select the behaviour
 that locally maximises the use of available resources to  yield the maximum \ac{QoS}. This may lead to a non-optimal solution, a limitation of our current work.

\textbf{Peer-awareness:} This refers to the ability to communicate with peers to collaboratively diagnose and respond to faults. In the proposed methodology,   components communicate via fault communication channels to effectively deal with faults. For example when a producer that is publishing sensor data needs to be restarted,  the consumer is informed that the producer is inactive and knows when it will be reactivated.

	\section{Proposed contract-based resiliency }
	\label{sec:propArch}
	
In this section, we describe a new methodology for developing distributed resilient architectures, called Contract-based Resiliency. It is motivated by component-based engineering which has been successfully used for large-scale system designs and the use of contract-based designs for explicitly pinpointing the
responsibilities and assumptions of a component~\cite{benveniste2015contracts}.   
In our approach, we 
 describe an application using a network of components  and use contracts to capture the behaviour of the components. 
  Finally, we detect failures (contract violation) and react (change contracts dynamically) to the disturbances, providing resiliency.
 
\begin{figure}[htbp]
	\includegraphics[width=0.90\columnwidth]{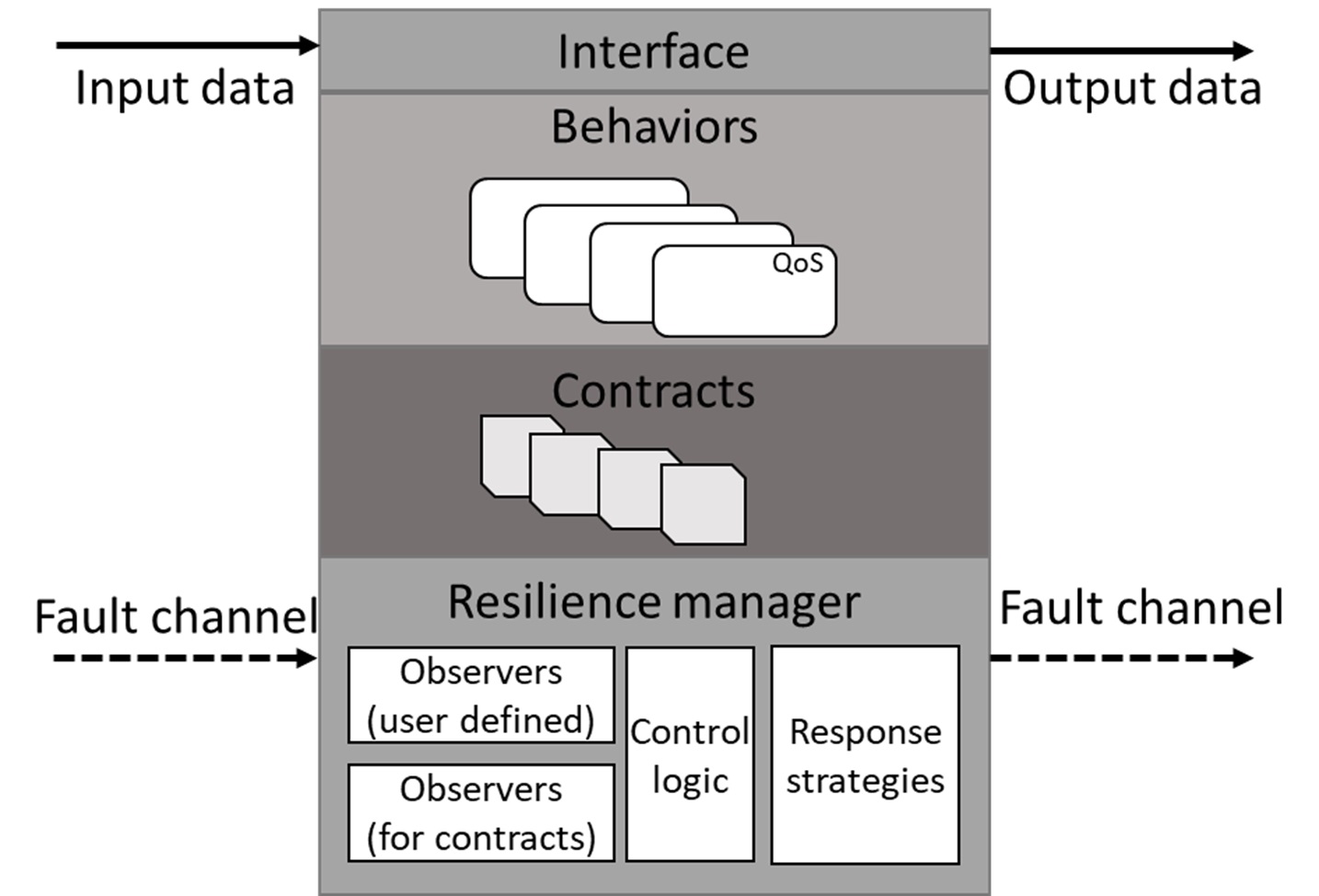}
	\centering
	\caption{\label{fig:component} Overview of the component model with the resilience manager. With respect to Fig.~\ref{fig:CILayers}, software components of application layer, computational and communication platforms of platform layer are considered as components.\vspace{-0.6cm}}
\end{figure}

\subsection{Component architecture}

A component is an open system that  receives inputs from the environment,
 executes a behaviour, and  generates output to the environment. 
The environment could be the collection of other components or the physical world. An overview of a component is presented in Fig.~\ref{fig:component}.

\begin{itemize}[leftmargin=*]
	\item \textbf{Interface:}
	It defines the Input/Output data channels of a component. Data is consumed through input interface, processed by the component, and output data is produced. 
	
	\item \textbf{Behaviours:} A component can be described using multiple behaviours.  Each behaviour is associated with a QoS. At runtime, the resilience manager selects the behaviour of the component.

	\item \textbf{Contracts:} A contract specifies assumptions on the behaviour of the environment, and guarantees about the behaviour of the component~\cite{benveniste2015contracts}. 
	At runtime, the resilience manager can switch between contracts to react to the disturbances in the system.
	
	\item \textbf{Resilience manager:} Detects faults (using Observers) and decides (control logic) the best course of action (response strategy). It also responds to the fault information from other components.

\end{itemize}

\subsection{Example application}

{\color{black}
Fig.~\ref{fig:egApp} presents the running example, 
called Conveyor Belt Block Pick-up (CBBP). 
The application relies on the timing information from sensors S1 and S2 and then
activate the robot at the right time to pick-up block from a moving conveyor belt.
}


\begin{figure}[hbtp]
	\includegraphics[scale=0.5]{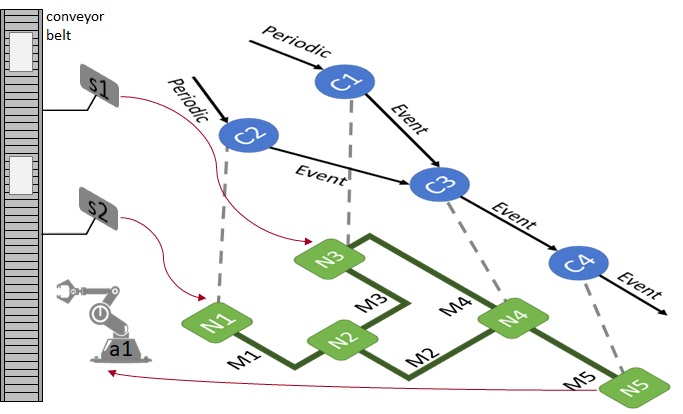}
	\centering
	\caption{\label{fig:egApp} Overview of the mapping between software components ($c_1,\dots,c_4$) of the application layer and the computational components ($n_1,\dots,n_5$) of the platform layer. Physical layer is not shown for brevity.
	\vspace{-0.6cm} }
\end{figure}

\textbf{Application graph:} The CBBP application is described as a set of connected components  in the form of a graph, called \emph{application graph}.
The nodes of the graph are components and the edges represent communication between components. See the graph in Fig.~\ref{fig:egApp} with components $c_1,c_2,c_3$ and $c_4$. 	Component~$c_1$  samples sensor ($S1$) data \emph{periodically} ($T_{samp}$).
\sid{It applies different signal processing techniques 
(described using different component Behaviours, see Fig.~\ref{fig:component})} to extract the time ($t_1$) when the centre of a {block} is aligned with sensor $S1$. When the {block} is detected, the timing information ($t_1$) is sent to component $c_3$. 
 Similarly,  component~$c_2$ periodically samples sensor~$S2$ to compute and send the timing information ($t_2$) to component $c_3$.
Based on the received timing information, $c_3$ computes the time ($t_3$) when the robot needs to be activated and then sends it to component~$c_4$.  Finally, the robot is activated to pick-up a block from the assembly line.

\textbf{Mapping between application and platform layers: } 
The platform layer comprises a set of computational platforms ($n_1, \dots, n_5$) 
and connecting communication platforms ($m_1,\dots,m_5$). 
The mapping of the components in an application graph to  computational platforms
is shown in Fig.~\ref{fig:egApp}. E.g., component~$c_1$ is mapped to  computational platform~$n_3$. Assuming that $c_1$ has four possible behaviours ($BEH \in \{beh_1^{c1}, beh_2^{c1}, beh_3^{c1}, beh_4^{c1}\}$), 

based on the mapping information we assume an execution \sid{time} cost function $EC:BEH \rightarrow \realN$ is given. 
Furthermore, we assume the communication between components is handled by a network infrastructure. We assume a communication \sid{time} cost function $CC:BEH \rightarrow \realN$ is given.
\vspace{-0.1cm}

\subsection{Generating contracts from timing deadlines (off-line)} 
{\color{black} Using the speed of the conveyor belt 
we can compute the time taken for a block to travel from  
sensor (S1) to the actuator (robot),} denoted as $\TsToCam$.  
We assume the decomposition of end-to-end timing constraints (e.g., 10 sec from S1 to robot) into 
deadlines for each component is given.
For the CBBP example, we assume the end-to-end latency
is decomposed as described by Eq.~\ref{equ:e2e}, where all terms are constants.
\begin{equation}
\label{equ:e2e}
 T_{s_1 \rightarrow c_1 } + 
 T_{c_1 \rightarrow c_3 } +
 T_{c_3 \rightarrow c_4 } + 
 T_{c_4 \rightarrow a_1 } \le  \TsToCam
\end{equation}
\vspace{-0.5cm}

\subsubsection{
	\label{sec:s1Contract}
	\textbf{Contracts to satisfy the timing constraint $\mathbf{T_{s_1 \rightarrow c_1 } } $}
}
Constant $T_{s_1 \rightarrow c_1 }$ refers to the time taken for $n_3$ to
read  sensor $s_1$ and 
 send the data to component $c_1$. Based on existing notation~\cite{benveniste2015contracts},  the first contract on $n_3$, 
  denoted by symbol $\conSym_{n_3}^{1}$ is as follows.
\begin{equation*}
\small
\conSym_{n_3}^{1}:\left\{
\begin{array}{@{}ll@{}}
inputs: & \text{$s_1$\_data $\in \realN$} \\
outputs: & \text{$c_1$\_data $\in \realN$} \\
assumptions: & \text{$\noAssumptions$} \\
guarantees: & \text{$c_1$\_data} = \text{$s_1$\_data}~\within~T_{s_1 \rightarrow c_1 }
\end{array}
\right.
\end{equation*} 

Furthermore, we know that the sensors must be sampled periodically every $T_{samp}$.
This requirement can be treated as a separate contract on $n_3$, $\conSym_{n_3}^{sampling}$. In this case, both contacts can be composed~\footnote{Rules of composition is a part of the future work. They can be based on existing work~\cite{benveniste2015contracts}. \vspace{-0.7cm}}. 

\begin{equation*}
\small
\conSym_{n_3}^{1^{\prime}}:\left\{
\begin{array}{@{}lll@{}}
inputs: & \text{$s_1$\_data $\in \realN$} \\
outputs: & \text{$c_1$\_data $\in \realN$} \\
assumptions: & \text{$\noAssumptions$} \\
guarantees: &\text{$c_1$\_data} = \text{$s_1$\_data}&\within~T_{s_1 \rightarrow c_1 }\\
&&\every~\text{$T_{samp}$}
\end{array}
\right.
\end{equation*}

\subsubsection{Contracts to satisfy the timing constraint $\mathbf{T_{c_1 \rightarrow c_3 }} $}
Constant $T_{c_1 \rightarrow c_3 }$ refers to the time taken for 
 $n_3$~to send the output data of $c_1$ to $n_4$ over a network
and $n_4$ to send the data to  $c_3$.

 Component~$c_1$ has four behaviours with different execution times on $n_3$, as described earlier. For each of the behaviours, we generate contracts. E.g., for behaviour $beh_1^{c1}$ the contract $\conSym_{n_3}^{c_1 beh_1}$ assumes nothing (represented using the symbol~$\noAssumptions$) and guarantees execution of $beh_1^{c1}$ \within $EC(beh_1^{c1})$.

 We do not assume the communication delay to be a constant. At runtime, communication interface on $n_3$ and the network controller engage to create a contract $\conSym_{n_3}^{c_1 n_4}$ that provides a time bound on the delivery of $c_1$'s data to $n_4$. 

 Based on an expected time ($T_{n_4 \rightarrow c_3 }$) for processing the received data on $n_4$ and sending it to $c_3$,  a single contract $\conSym_{n_4}^{c_1 c_3}$ is created where it assumes nothing and guarantees the delivery within $T_{n_4 \rightarrow c_3 }$.

 Finally,  to monitor the  timing constraint $\mathbf{T_{c_1 \rightarrow c_3 }}$, observer $\obsSym^{1}_{n_4}$ executes on $n_4$ to monitor the timestamps on the data from $c_1$ to $c_3$. Component $n_4$ was chosen since $c_3$ is mapped to $n_4$.
 Note that if the observer was to execute on any other platform component such as $n_5$, then timing information from $n_4$ need to be  sent to $n_5$. This unnecessarily increases the fault detection time of the architecture.

\subsection{Runtime reconfiguration}
 
\textbf{1. Application-level reconfiguration:}
The state of the application is described by an ordered set of contracts currently
selected by the components to execute their behaviour. 
For the running example, the ordered set of components
is $[c_1,c_2,c_3,c_4]$. The initial state of the application can be $[\conSym_{n_3}^{c_1 beh_1},\conSym_{n_1}^{c_2 beh_1}, \conSym_{n_4}^{c_3 beh_2}, \conSym_{n_5}^{c_4 beh_1}] $.  Any \emph{application-level reconfiguration} is a transition from one application state to another.
This state transition occurs when a fault is detected (violation of a contract) and a new contract is adopted by a software component.

\textbf{2. Platform-level reconfiguration:} As described earlier, the platform layer consists of computational and communicational platforms. Once again, the state of the platform can be described as an ordered set of contracts e.g., $[\conSym_{n_3}^{c_1 n_3},\conSym_{n_4}^{c_1 c_3},\dots]$. 
Any \emph{platform-level reconfiguration} is a transition from one platform state to another. E.g., due to high network traffic the contract $\conSym_{n_3}^{c_1 n_3}$ needed to be re-negotiated  to $\conSym_{n_3}^{c_1 n_3 \prime}$, representing a reconfiguration at platform level. Note, this may or may not lead to an application level reconfiguration.
		
	\section{Fault-detection and Response}
	\label{sec:conclusions}
	In the proposed resilience architecture, fault detection is
tightly coupled with the contracts.
A fault detection (violation of contracts monitored using observers)  and response (change in contract by the resilience manger) will be managed efficiently.
In the following, we discuss some of the faults experienced in distributed systems~\cite{Kola2005}.

\subsection{Hanging processes}
\label{sec:hangingProc}
In this case, execution time of a process is longer than normal. 
For our example, component~$c_1$ may take longer than expected. 
The time constraints are captured using the contract $\conSym_{n_3}^{1^{\prime}}$ in Sec.~\ref{sec:s1Contract}. It states
that the process must be completed within the duration of $T_{s_1 \rightarrow c_1 }$ and the process must be executed every  $T_{samp}$.

\textbf{Fault detection technique:}
The contract $\conSym_{n_3}^{1^{\prime}}$ is  associated with an observer. The observer $\obsSym^{1^{\prime}}_{n3}$ is implemented using timed automaton~\cite{Alur:1994}, see Figure~\ref{fig:obsAsTA}.
\vspace{-0.6cm}
\begin{figure}[hbtp]
	\begin{tikzpicture}[->,>=stealth',shorten >=1pt,auto,
node distance=3cm,
semithick,scale=0.8, transform shape]
\tikzstyle{every state}=[rectangle,rounded corners,
minimum height = 1.5cm, text width=1.0cm, text centered, fill=blue!20,draw=none,text=black, draw,line width=0.3mm]
\tikzstyle{line} = [draw, -latex']

\node[state, fill=backgroundFourth,
label={[shift={(0.0,-2)}]\small $ \mathbf{Idle}$ }]
(IDLE)  {$\dot{t}= 1 $};

\path[<-, dashed] (IDLE.110) edge node[below, align=left, shift={(0.5,1)}] {
	\footnotesize initial \\
	\footnotesize $\begin{matrix}
    {t=0}{}
	\end{matrix}$
} ++(0cm,1cm);

\node[state, fill=backgroundSeventh,
label={[shift={(0,0)}]}, 
label={[shift={(0,-2)}]\small $\mathbf{Process}$   }]
(PROC) [right of=IDLE] {$\dot{t}=1$};

\node[state, fill=red!20,
label={[shift={(0,0)}]$ $}, 
label={[shift={(0,-2)}]\small $\mathbf{Error 1}$   }]
(E1) [left of=IDLE] {$\dot{t}=1$};

\node[state, fill=red!20,
label={[shift={(0,0)}]$ $}, 
label={[shift={(0,-2)}]\small $\mathbf{Error 2}$   }]
(E2) [right of=PROC] {$\dot{t}=1$};





\path[->] (IDLE.35) edge node[align=center] {
	$\frac{ start! }{t=0}$ \\   
} (PROC.145);

\path[->] (PROC.-145) edge node[align=center] {
	$\frac{ end! }{t=0}$ \\   
} (IDLE.-35);

\path[->] (IDLE) edge node[below, align=left] {
	$\frac{t > T_{samp}}{t=0, err(1)!}$
} (E1);

\path[->] (PROC) edge node[below, align=left] {
	$\frac{t > T_{s_1 \rightarrow c_1}}{t=0, err(2)!}$
} (E2);

\draw (E1) to[out=90, in=120] (IDLE) node[above right =0.2cm of E1] {$reset!$};

\draw (E2) to[out=130, in=70] (IDLE) node[above left =0.7cm and 0.1cm of E2 ] {$reset!$};

\end{tikzpicture}

	\vspace{-0.5cm}
	\caption{\label{fig:obsAsTA}An observer implemented using timed automata. It detects if the process deadline ($T_{s_1 \rightarrow c_1 }$) or sampling rate ($T_{samp}$) are not satisfied.  
		\vspace{-0.3cm}}
\end{figure}
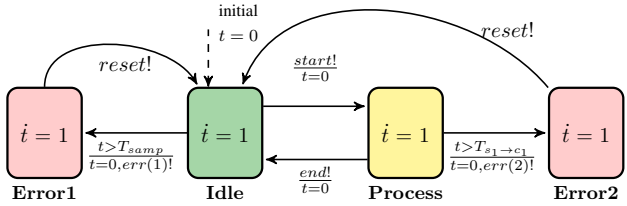

\textbf{Response to the fault:}
\sid{The observer failure is detected by the control logic of the resilience manager.
As a response strategy, the component~$c_1$ is restarted
and the resilience manager of component~$c_3$ is notified about the type of fault and the expected recovery time
via the fault channel, see Fig.~\ref{fig:component}. 

Based on the recovery time of a producer (e.g., $c_1$), the control logic of the consumer (e.g., $c_3$) may select different response strategies.
 For example, $c_3$ can reduce its execution time by selecting a  different behaviour. 
In this case, based on the  recovery time   $beh_3^{c1} $  can be selected instead of $beh_1^{c2}$.
Of course, this reduces the quality of service.
This response is achieved by switching the contracts at runtime from $\conSym_{n_3}^{c_1 beh_1}$ to $\conSym_{n_3}^{c_1 beh_2}$, resulting in an application-level reconfiguration.}

\subsection{Network outages}
\sid{In this case, communication link  has failed.
For our example, consider the connection between $n_3$ and $n_4$.}

\textbf{Fault detection technique:}  
\sid{Using heartbeat signals component-level resilience managers of $n_3$ and $n_4$ periodically detect the status of the connection.}

\textbf{Response to the fault:} 
\sid{Observers fail when the heartbeat signals are not received by the resilience managers of $n_3$ and $n_4$.
This triggers the control logic which can selects a response strategy such as switching from a wired to a wireless connection.
Furthermore, if the component-level resilience managers were unable to re-establish the link,
they would inform a layer-level resilience manager such as an SDN controller to find an alternative route.}


	\section{Conclusions \& Future work }
	\label{sec:conclusions}
	
We have presented a contract-based methodology 
for enabling resilient cyber-infrastructure for Industry~4.0. 
Applications are described as a set of modular components
that are distributed over a network. Contracts are used for
describing the component's interaction with other components (and or across layers).
Finally, the terms of the contract are monitored using observers.
We detect failures (contract violation) and react (change contracts dynamically) to the disturbances, for providing resiliency.

In future,  we plan to develop a multi-dimensional resilience metric to  evaluate resilience with respect to different performance indicators such as security, safety, throughput, recovery time, etc. 
\sid{Furthermore, we plan to support parametrized
	contracts to automatically respond to  faults across different components.
	E.g.,  if the robot is unable to meet the deadline to successfully pick-up blocks from the moving conveyor as a response the motor speed can be reduced (update contract w.r.t. conveyor speed). This idea is along  our vision to support plug-n-produce which requires dynamic reconfiguration.}


		\bibliographystyle{ieeetr}
	\bibliography{references}

\begin{thebibliography}{10}

\bibitem{LEE201518}
L.~Jay, B.~Behrad, and K.~Hung-An, ``{A Cyber-Physical Systems architecture for
  Industry 4.0-based manufacturing systems},'' {\em Manufacturing Letters},
  vol.~3, pp.~18--23, Jan. 2015.

\bibitem{Vyatkin2017}
W.~Dai, V.~N. Dubinin, J.~H. Christensen, V.~Vyatkin, and X.~Guan, ``{Toward
  Self-Manageable and Adaptive Industrial Cyber-Physical Systems With
  Knowledge-Driven Autonomic Service Management},'' {\em IEEE Transactions on
  Industrial Informatics}, vol.~13, pp.~725--736, April 2017.

\bibitem{Sztipanovits2012}
J.~Sztipanovits and et~al., ``{Toward a Science of Cyber-Physical System
  Integration},'' {\em Proceedings of the IEEE}, vol.~100, pp.~29--44, Jan.
  2012.

\bibitem{Strigini2012}
L.~Strigini, ``{Fault Tolerance and Resilience: Meanings, Measures and
  Assessment},'' in {\em {Resilience Assessment and Evaluation of Computing
  Systems}} (K.~Wolter and et~al., eds.), pp.~3--24, 2012.

\bibitem{Eisele2017}
E.~Scott, I.~Madari, A.~Dubey, and G.~Karsai, ``{RIAPS:Resilient Information
  Architecture Platform for Decentralized Smart Systems},'' in {\em
  International Symposium on Real-time Computing}, IEEE, May 2017.

\bibitem{OliverH2017}
H.~Oliver, I.~Haris, S.~Muhammad, and G.~Radu, ``{A Self-Healing Framework for
  Building Resilient Cyber-Physical Systems},'' in {\em Proceedings of the Int.
  Symposium on Real-time Distributed Computing}, IEEE, 2017.

\bibitem{Valls2013}
M.~G. Valls, I.~R. Lopez, and L.~F. Villar, ``{iLAND: An Enhanced Middleware
  for Real-Time Reconfiguration of Service Oriented Distributed Real-Time
  Systems},'' {\em IEEE Transactions on Industrial Informatics}, vol.~9,
  pp.~228--236, Feb 2013.

\bibitem{Lepuschitz2011}
W.~Lepuschitz, A.~Zoitl, M.~Vallée, and M.~Merdan, ``{Toward
  Self-Reconfiguration of Manufacturing Systems Using Automation Agents},''
  {\em IEEE Transactions on Systems, Man, and Cybernetics, Part C (Applications
  and Reviews)}, vol.~41, pp.~52--69, Jan 2011.

\bibitem{Denker2012}
G.~Denker and et~al., ``Resilient dependable cyber-physical systems: a
  middleware perspective,'' {\em Journal of Internet Services and
  Applications}, vol.~3, pp.~41--49, May 2012.

\bibitem{benveniste2015contracts}
A.~Benveniste and et~al., ``{Contracts for Systems Design: Theory [Research
  Report] RR-8759},'' tech. rep., 2015.

\bibitem{Kola2005}
G.~Kola, T.~Kosar, and M.~Livny, ``{Faults in Large Distributed Systems and
  What We Can Do About Them},'' in {\em Proc. of the Int. Euro-Par Conference
  on Parallel Processing}, pp.~442--453, Springer-Verlag, 2005.

\bibitem{Alur:1994}
R.~Alur and D.~L. Dill, ``{A Theory of Timed Automata},'' {\em Theor. Comput.
  Sci.}, vol.~126, pp.~183--235, Apr. 1994.

\end{thebibliography}
	\cleardoublepage







	
\end{document}